\documentclass{amsart}
\usepackage{amssymb, array, verbatim, amscd}
\usepackage{amsmath, amsfonts,amsthm,graphicx}

\newtheorem{theorem}{Theorem}[section]
\newtheorem{corollary}[theorem]{Corollary}

\newtheorem{definition}[theorem]{Definition}
\newtheorem{example}[theorem]{Example}

\newcommand{\Z}{{\mathbb Z}}
\newcommand{\R}{\mathbb R}

\begin{document}
\vspace*{8mm}
\title[Probabilistic Regulatory Networks] { Probabilistic Regulatory Networks: modeling genetic networks}
\author{Mar\'{\i}a A. Avino-Diaz and Oscar Moreno  }
\address { Department of Mathematic-Physics, Cayey, mavino@uprr.pr\\
Department of Mathematics and Computer Sciences, Rio Piedras, moreno@uprr.pr\\
University of Puerto Rico}
\maketitle
\begin{abstract}  We describe here the new concept of
$\epsilon$-Homomorphisms of Probabilistic Regulatory Gene
 Networks(PRN). The $\epsilon$-homomorphisms are special mappings between two probabilistic networks, that
 consider the algebraic action of the iteration of functions and
 the probabilistic dynamic of the two networks. It is proved here that the class of PRN, together
with the homomorphisms, form a category with products and
coproducts. Projections are special homomorphisms, induced by
invariant subnetworks. Here, it is proved that an
$\epsilon$-homomorphism for $0<\epsilon<1$ produces simultaneous
Markov Chains in both networks, that permit to introduce the
concepts of $\epsilon$-isomorphism of Markov Chains, and similar
networks.
\end{abstract}

\section*{ Introduction}
 We can understand the complex interactions of genes using
simplified models, such as discrete or continuous models of genes.
Developing computational tools permits description of gene functions
and understanding the mechanism of regulation \cite{D2005,50}. We
focus our attention in the discrete structure of genetic regulatory
networks instead of continuous models. Probabilistic Gene Regulatory
Network (PRN) is a natural generalization of the Probabilistic
Boolean Network (PBN) model introduced  in \cite{SDZ}, and
\cite{A1}. This model have $n$ functions defined over a finite set
$X$ to itself, with probabilities assigned to these functions. We
present here the ideas of $\epsilon$-similar networks, and
isomorphism of Markov Chain. $\epsilon$-homomorphisms are  used to
describe subnetworks and similar networks, because they transform
the discrete structure of one network to another, and the
probability distributions of the networks are  enough close, using a
preestablished $0<\epsilon<1 $ as a distance between the
probabilities.
\section{Preliminaries}
\vspace{0.2in} \noindent{\bf Probabilistic Regulatory Networks}
 A Probabilistic Gene Regulatory Network (PRN)  (or a Probabilistic
 Dynamical Systems)\cite{A1} is
a triple $\mathcal{X}=(X,F,C)$ where $X$ is a finite set and
$F=\{f_1, \ldots , f_n\}$ is a set of functions from $X$ into
itself, with a list $C=(c_1, \ldots ,c_n)$ of selection
probabilities, where $c_i=p(f_i)$, \cite{A1}
 We associate with each PRN  a weighted digraph, whose vertices are the elements of $X$, and
if $u,v\in X$, there is an arrow going from $u$ to $v$ for each
function $f_i$ such that $f_i(u)=v$, and the probability $c_i$ is
assigned to this arrow. This weighted digraph will be called the
state space of $\mathcal{X}$. In this paper, we use the notation PRN
for one or more networks. If $X=X_1\times\cdots X_n$ is the product
of $n$ sets of possible values of the variables, then with the
vector function $f=(f^1,\cdots,f^n)$ we associate a digraph
$\Gamma$, called dependency graph, with vertex set $\{1,\dots , n
\}$. There is a directed edge from $i$ to $j$ if $xi$ appears in the
component function $f^j$. For a PRN, we have a dependency graph
(dep-graph) for each function, then we superpose all the dep-graph
and that is the low level digraph of our PRN \cite{SDZ}

 \vspace{0.2in} \noindent{\bf Example.} Suppose we have two genes with two values that we denote as usual $\{0,1\}$, that is this PRN is a
very simple  PBN. The set of boolean functions $F$ is the following:
\[F=\{ f_1(x1,x2)=(x1,0), f_2(x1,x2)=(1,x2),\] \[
f_3(x1,x2)=(1,0), f_4(x1,x2)=(x1x2,x2)\},\] and the probabilities
are $\{.21,.22,.34,.23\}$. Therefore,  the PBN $\mathcal{X}=(X,F,C)$
has the following state space,  dependency graph, and  transition
matrix.
\[\begin{array}{c}
\overset{.66}{\circlearrowright}(0,0)\leftarrow ^{.21} (0,1)\circlearrowleft^{.23}\\
^{.34}\downarrow \uparrow^{.23} \hbox{  }\swarrow_{ .34}\downarrow^{.22}\\
\overset{.77}{\circlearrowright}(1,0)\overset{.55}{\longleftarrow}
(1,1)\circlearrowleft^{.45} \\
\hbox{ State space}\cr
\end{array}
\hspace{1.1in}\begin{array}{c}
{\curvearrowleft}\hspace{.6in} {\curvearrowleft}\\
 \underline{\overline{|\textmd{gene \ 1}|}}
\hspace{.1in}{\longleftarrow} \hspace{.1in}
\underline{\overline{|\textmd{gene \ 2}|}} \\
\hbox{ Dependency graph}\\
\hbox{ of genes x1 and x2}\cr
\end{array}\]
\[ T=\left[\begin{array}{cccc} .66&0&.34&0\\
.21&.23&.34&.22\\
.23&0&.77&0\\
0&0&.55&.45\cr
\end{array}\right]\]

\vspace{0.2in} \noindent{\bf $\epsilon$-Homomorphisms of PRN.} If
$C$ is a set of selection probabilities we denote by $\chi$ the
characteristic function over $C$. That is $\chi:C\cup\{0\}
\rightarrow \{0,1\}$ such that $\chi(c)=1$, if $c\ne 0$ and
$\chi(0)=0$. Let $\mathcal{X}_1=(X_1,F=(f_i)_{i=1}^n ,C)$ and
$\mathcal{X}_2=(X_2,G=(g_j)_{j=1}^m, D)$ be two PRN.  A map
$\phi:X_1\rightarrow X_2$ is  an \textbf{$\epsilon$-homomorphism}
from $\mathcal{X}_1$ to $ \mathcal{X}_2$,  if for a fixed real
number $0\le \epsilon<1$, and for all $f_i$ there exists a $g_j$,
such that for all $u$, $v$ in $\mathcal{X}_1$,
\begin{center}(1) $\phi \circ f_i=g_j\circ \phi;$ (2) $max_{u,v}|c_{f_i}(u,v)-d_{g_j}(\phi(u),\phi(v))|\le
\epsilon$, and \\(3) $\chi(d_{g_j}(\phi(u),\phi(v)))\geq
\chi(c_{f_i}(u,v)).$
 \end{center}
If $\phi:X_1\rightarrow X_2$ is a bijective map, and
$d_{g_j}(\phi(u),\phi(v))=c_{f_i}(u,v)$, for all $f_i$, $g_j$, $u$,
and $v$ in $\mathcal{X}_1$; then $\phi$  is an isomorphism.

If we denote by $p(u,v)=\sum_{f_i}c_{f_i}(u,v)$ and
$p(\phi(u),\phi(v))=\sum_{g_j}d_{g_j}(\phi(u),\phi(v))$, then
condition (2) implies that $|p(u,v)-p(\phi(u),\phi(v))|\leq k
\epsilon$, where $k$ is the maximum number of functions going from
one state  to another in the network. So, if $T_1$ denote the
transition matrix of $\mathcal{X}_1$, and the entry $(u,v)$ of $T_1$
is $p(u,v)$ then the third condition implies that:
$max_{u,v}|({T_1})_{u,v}-({T_2})_{\phi(u),\phi(v)}|\leq k\epsilon$,
 for all possible  $u$ and $v$ in $\mathcal{X}_1$.

\section{Isomorphism of Markov Chains, $\epsilon$-Similar Networks }

 Two PRN  are $\epsilon$-similar if  there exists a bijective homomorphism  $\phi$
  between them, such that $\phi ^{-1}$ is also an homomorphism.
  Observe that $\phi$ and $\phi^{-1}$ have the same $\epsilon$. When
  two PRN are $\epsilon$-similar, the two transition matrices have
  the a similar distribution of probabilities.
\begin{theorem}\label{teo}
If $\phi:\mathcal{X}_1\rightarrow\mathcal{X}_2$, and $\phi^{-1}$ are
 bijective $\epsilon$-homomorphisms, then
\begin{center}max$|(c_{f^m}(u,f^m(u))-d_{g^m}(\phi(u),g^m(\phi(u)))|\leq
m\epsilon$,
\end{center}
 for all $m>2$;  $u$, $v$,  in $\mathcal{X}_1$.

\end{theorem}
\begin{proof}
If $\chi(c_{f}(u,f(u)))=1$, then
$\chi(d_{g}(\phi(u),\phi(f(u))))=1$, because $\phi$ and $\phi ^{-1}$
are bijective homomorphisms.
 By definition of $\epsilon$-homomorphism, $g(\phi(u))=\phi(f(u))$. Then for $m= 2$,
 and by the Chapman-Kolmogorov equation \cite{KCS}, we
have the following:
\[|c_{f^2}(u,f^2(u))-d_{g^2}(\phi(u),g^2(\phi(u)))|=\]
\[|c_f(u,f(u))c_f(f(u),f^2(u))-d_g(\phi(u),g(\phi(u)))d_g(g(\phi(u)),g^2(\phi(u)))|=\]
\[|c_f(u,f(u))c_f(f(u),f^2(u))-d_g(\phi(u),\phi(f(u)))d_g(\phi(f(u)),\phi(f^2(u)))|\leq\]
By condition (2) in definition of homomorphism, we have\[\leq
|c_f(f(u),f^2(u))|\epsilon+|d_g(\phi(u),\phi(f(u)))|\epsilon\leq
2\epsilon.\]

Then we  proved that
$|c_{f^2}(u,f^2(u))-d_{g^2}(\phi(u),g^2(\phi(u)))|\leq 2 \epsilon$.

 Using
this property, and mathematical induction  over $m$, we can conclude
that our claim holds.
\end{proof}
\begin{corollary}If $\phi:\mathcal{X}_1\rightarrow\mathcal{X}_2$, and $\phi^{-1}$ are
 bijective $\epsilon$-homomorphisms, then the transition matrices
$T_1$ and $T_2$ satisfy the condition:
\begin{itemize}

\item[1.] $\chi(T_1^m)_{u,v}=\chi(T_2^m)_{\hat{u},\hat{v}}$,

\item[2.] $\sum_{i=1}^n ((T_1^m)_{u,v}-(T_2^m)_{\hat{u},\hat{v}})=0$,
\end{itemize}
for all $m$, $\hat{u}=\phi(u)$,  and $\hat{u}=\phi(u)$.
\end{corollary}
An $\epsilon$-homomorphism between two PRN determines a
correspondence between the Markov Chains of these two  networks.
Here, we introduce the concept of two similar Time Discrete Markov
Chain (TDMC).
\begin{definition} Two TDMC of the same size $n\times n$:
$\{T_1,\ T_1^2, \ T_1^3,\  \ldots\}$, and $\{T_2,\ T_2^2, \ T_2^3, \
\ldots\}$ are $\epsilon$-similar  or $\epsilon$-isomorphic if there
exists an $\epsilon \in \R$  small enough, such that
$T_1^m-T_2^m=(t_{ij})_{n\times n}$ satisfies that
\begin{itemize}
\item[(1)]  $|t_{ij}|<\epsilon$, and $\sum_{i=1}^n t_{ij}=0$,
\item[(2)] $\chi(T_1^m)_{ij}=\chi(T_2^m)_{ij}$,  for all $m$, where $\chi$ is the characteristic function.
\end{itemize}
 That is, these two TDMC simulated the dynamic of two $\epsilon$-similar networks.
\end{definition}

\begin{example}
\end{example}The networks  with dynamic $T_1$ and $T_2$ are $.005$-similar. In fact \[  T_1=\left[\begin{array}{cccc} 0&.549&.451&0\\
0&.338&0&.662\\
.111&.445&.444&0\\
0&.013&0&.987\cr
\end{array}
\right] T_2=\left[\begin{array}{cccc} 0&.544&.456&0\\
0&.337&0&.663\\
.113&.448&.439&0\\
0&.011&0&.989\cr
\end{array}
\right]\] Observe that,
\[T_1-T_2=\left[\begin{array}{cccc} 0&.005&-.005&0\\
0&.001&0&-.001\\
-.002&-.003&.005&0\\
0&.002&0&-.002\cr
\end{array}
\right]\] As a consequence, we obtain
$max|(T_1)_{ij}-(T_2)_{ij}|\leq .005$, and both dynamics are
.005-isomorphic. The steady state of $T_1$ is
$\pi_1=(0,.01926,0,.98074)$, and the steady state of $T_2$ is
$\pi_2=(0,.01632,0,.98368)$, \cite{KCS}. We can see that
$|\pi_1-\pi_2|=max_i|\pi_1(i)-\pi_\phi(i)|<.004$. Additionally, we
have
\[{T_1}^2-{T_2}^2=\left[\begin{array}{cccc}
-.001467&-.00136&.00006&.00277\\
0&.00 199&0&-.00 199\\
-.00232&-.00019&.00295&-.00243\\
0&.002639& 0&-.00 263 \cr
\end{array}
\right],\] therefore $max|(T_1^2)_{ij}-(T_2^2)_{ij}|\leq .003$.
\[T_1^3-{T_2}^3=\left[\begin{array}{cccc}
-.000394&-.00044&.00011&.00073\\
0&.002525&0&-.00253\\
-.000161&.00156&.00213&-.00353\\
0&.002843&0&-.002843\cr
\end{array}
\right],\] and $max|(T_1^3)_{ij}-(T_2^3)_{ij}|\leq .004$. In the
above example, the TDMC generated by $T$ and $T_2$ are
$.005$-similar, and the networks simulated by them are
$.005$-similar.
\section{ The category of Probabilistic Regulatory Networks, and mathematical background}
For a $\epsilon \in \R$ small enough, we have the following theorem.
\begin{theorem}
  If $\phi_1:\mathcal{X}_\rightarrow \mathcal{X}_2$, and $\phi_2:\mathcal{X}_2\rightarrow
\mathcal{X}_3$ are $\epsilon_i$-homomorphisms, for $i=1,2$. Then
$\phi=\phi_2\circ \phi_1:\mathcal{X}_1\rightarrow \mathcal{X}_3$ is
an $\epsilon$-homomorphism. Therefore the Probabilistic Regulatory
Networks with the $\epsilon$-homomorphisms of PRN form the category
\textbf{PRN}.
\end{theorem}
\begin{proof}
The Probabilistic Regulatory Networks with the PRN homomorphisms  is
a category if: the composition is an homomorphism, and satisfy the
associativity law; and there exists an identity homomorphism for
each PRN.

  (1) Let $\phi_1:\mathcal{X}_1\rightarrow \mathcal{X}_2$ be an $\epsilon_1$-homomorphism, and let
$\phi_2:\mathcal{X}_2\rightarrow \mathcal{X}_3$ be an
$\epsilon_2$-homomorphism. If $h\in \mathcal{X}_3$, $g\in
\mathcal{X}_2$ and $f\in \mathcal{X}_1$ are functions in each PRN,
and such that $\phi_1 \circ f=g\circ \phi_1$ and $\phi_2 \circ g=h
\circ \phi_2$, then we will prove that: $\phi \circ f=h \circ \phi.$
In fact,
 \[(\phi_2 \circ \phi_1) \circ f=\phi_2 \circ (\phi_1 \circ f)=\phi_2 \circ (g
 \circ \phi_1)=(\phi_2 \circ  g)\circ \phi_1=\]
 \[(h \circ \phi_2)\circ \phi_1=h \circ (\phi_2\circ \phi_1).\]
\noindent

(2) To verify the second condition for $\epsilon$-homomorphism, we
do the following. If $c_f(\phi(u),\phi(v))\ne 1$, with $u,\
v=f(u)\in X_1$, for some $f\in \mathcal{X}_2$,then we will  prove
that there exists an $\epsilon<1$ such that
\[|c_f(u,v)-t_h(\phi(u),\phi(v))|<\epsilon.\]
by part (1). We denote by  $\hat{u}=\phi_1(u)$, $\hat{v}=\phi_1(v)$.
\[|c_f(u,v)-d_g(\phi_1(u),\phi_1(v))+d_g(\phi_1(u),\phi_1(v))-t_h(\phi_2(\hat{u}),\phi_2(\hat{v}))|\leq\]
\[|c_f(u,v)-d_g\phi_1(u),\phi_1(v)|+|d_g(\phi_1(u),\phi_1(v))-t_h(\phi_2(\hat{u}),\phi_2(\hat{v}))|\leq\]
Therefore our claim holds,
$|c_f(u,v)-t_h(\phi(u),\phi(v))|<\epsilon_1+\epsilon_2.$

(3) We want to prove that
$\chi(t_h(\phi(u),\phi(v)))\ge{\chi(c_f(u,v))}.$ Suppose that
${\chi(c_f(u,v))}=1$. Then, since $\phi_1$ is an homomorphism of
PRN, we have that
\[\chi(d_g (\phi_1(u),\phi_1(v)))\ge {\chi (c_f(u,v))=1
}\]  Since $\phi_2$ is an homomorphism of PRN, we obtain that
\[\chi(t_h(\phi(u),
\phi(v)))=\chi(t_h(\phi_2(\phi_1(u)), \phi_2(\phi_1(v)))) \ge
\chi(c_f(\phi_1(u),(\phi_1(v))=1.\] Therefore we have that $\chi
(t_h(\phi_2(\phi_1(u)), \phi_2(\phi_1(v))))=1.$

Then the composition of two PRN-homomorphisms  is an homomorphism.

The associativity and identity laws are easily checked, then our
claim holds, and \textbf{PRN} is a category.
\end{proof}
For  proofs of the following theorems see \cite{A}
\begin{theorem}\label{product}
Let $\mathcal{X}_1\times \mathcal{X}_2=(X_1\times X_2,H,E)$ be a
product of PRN $\mathcal{X}_1=(X_1,F,C)$ and
$\mathcal{X}_2=(X_2,G,D)$. If $\delta_i:X \rightarrow X_i$ are two
PRN-homomorphisms, then there exists an homomorphism $\delta:
X\rightarrow X_1\times X_2$, such that $\phi_i\circ \delta=\delta_i$
for $i=1,2$. That is, the following diagram commutes
\[\begin{array}{c}
  \hspace{.05in} X_1\times X_2 \hspace{.05in}\\
 \overset{\phi_1}{ \swarrow}\hspace{.1in}\overset{\delta}{\uparrow}\hspace{.1in}\overset{\phi_2}{\searrow}\\
  X_1\overset{\delta_1}{\longleftarrow} X \overset{\delta_2}{\longrightarrow} X_2\cr
  \end{array}\]
  This homomorphism is unique.
\end{theorem}
\begin{theorem}
Let $\mathcal{X}_1\oplus \mathcal{X}_2=(X_1\times X_2,H,E)$ be a
product of PRN $\mathcal{X}_1=(X_1,F,C)$ and
$\mathcal{X}_2=(X_2,G,D)$. If $\gamma_i:X_i \rightarrow X$ are two
PRN-homomorphisms, then there exists an homomorphism $\gamma:
X_1\oplus X_2\rightarrow X $, such that $\gamma \circ
\iota_i=\gamma_i$ for $i=1,2$. That is, the following diagram
commutes
\[\begin{array}{c}
  \hspace{.05in} X_1\oplus X_2 \hspace{.05in}\\
 \overset{\iota_1}{ \nearrow}\hspace{.1in}\overset{\gamma}{\downarrow}\hspace{.1in}\overset{\iota_2}{\nwarrow}\\
  X_1\overset{\gamma_1}{\longrightarrow} X \overset{\gamma_2}{\longleftarrow} X_2\cr
  \end{array}\]
  This homomorphism is unique.
\end{theorem}
\section{Subnetworks}
A subnetwork  $Y\subseteq X$  of $\mathcal{X}=(X,F,C)$ is an
\textbf{invariant subnetwork or a sub-PRN} of $\mathcal{X}$ if
 $f_i(u)\in Y$ for all $u\in Y$, and  $f_i\in F$. Sub-PRNs
 are sections of a PRN, where there aren't arrows going out.
  The complete network $X$, and any cyclic state with
  probability 1, are sub-PRNs. An invariant subnetwork is irreducible if doesn't have a
proper invariant subnetwork.
 \emph{An endomorphism is a projection if $\pi^2=\pi$. }
\begin{theorem}\label{theo}
If there  exists a projection from $\mathcal{X}$ to a subnetwork
$\mathcal{Y}$ then $\mathcal{Y}$ is  an invariant subnetwork of
$\mathcal{X}$.
\end{theorem}
\begin{proof}
Suppose that there exists a projection $\pi:X\rightarrow Y$. If
$y\in Y$,  by definition of projection $\pi(y)=y$,  and
$f_i(\pi(y))=\pi(g_j(y))$. Therefore all  arrows in the subnetwork
$Y$ are going inside $Y$, and the network is invariant.
\end{proof}
\subsection{ Constructing a PRN with real data}
Here we developed a method to construct a PRN. In this case, we
suppose that the information given by the experiment is a dependency
graph and a time series data, see Figure 1, and Table 1.
\begin{table}[!ht]
\begin{center}
\begin{tabular}{|r|c|c|c|c|}
\hline $f_1$-data &3& 6 & 9  & 12  \\ \hline
x1 &2&2&2&2\\
\hline x2&1& 0&0&0\\ \hline x3 &0& 1&0 &0\\\hline
\end{tabular}
\begin{tabular}{|r|c|c|c|c|}
\hline $f_2$-data &3& 6 & 9  & 12 \\ \hline
x1 &1&1&1&1\\
\hline x2&0& 1&0&0\\ \hline x3 &1& 0&0 &0\\\hline
\end{tabular}
\begin{tabular}{|r|c|c|c|c|}
\hline $f_3$-data &3& 6 & 9  & 12  \\ \hline
x1 &2&0&0&0\\
\hline x2&0& 0&0&0\\ \hline x3 &1& 1&1 &1\\\hline
\end{tabular}
\caption{Time series data} \label{tbl1}
\end{center}
\end{table}

Additionally, we know that this information is noisy, and the first
gene has three values, meanwhile the other two genes take only two
$\{0,1\}$, so $X=\{0,1,2\}\times\{0,1\}^2$.

To determine the partially defined functions: $f_1,\ f_2, \ f_3$
over the finite field with $3$ elements  $\Z_3$, we use the
algorithm introduced in \cite{AGM}. That is: the first variable
$x1\in \Z_3$, meanwhile the other two genes $x_2,$ and $x_3$ are in
$\Z_2$.

For example with the first function $f_1=(f_{11},f_{12},f_{13})$ we
do the following. We represent the functions with polynomials over
the variables given by the dependency graph, and the operations $+$
and $\cdot$ are the usual in  the finite field $\Z_3$. Then, the
second component function
\[f_{12}(x1,x2,x3)=a+bx1+cx2+dx3+ex1x2+gx1x3+hx2x3+tx1x2x3\]  takes
the following table of values. $\begin{tabular}{|r|c|c|c|c|} \hline
$f_{12}$ (mod 2)  &1& 0 & 0  & 0 \\ \hline
x1 &2&2&2&2\\
\hline x2&1& 0&0&0\\ \hline x3 &0& 1&0 &0\\\hline
\end{tabular}$ . Evaluating, we obtain the following linear
system, where ``=" means congruence (mod 2):
\[\left\{\begin{array}{rl}
a+2b+c+2e&=0 \\
a+2b+d+2g&=0\\
a+2b&=0\cr
\end{array}\right..\]
Then reducing modulo 2, we have $a=c=d=0$, and $b,\ e,\ g,\ h,\ t,$
are free variables. So, one of the solution is
$f_{12}=x1(1+x2+x3+x2x3),$ (mod 2). Using this method, we obtain the
following functions:
\[\begin{array}{rl}
f_1(x1,x2,x3)&=(x_1,x1(1+x2+x3+x2x3),x_2),\\
 f_2 (x_1,x_2,x_3)&=(x1,x3,0), \\
f_3(x_1,x_2,x_3)&=(x1x2,x2,x3),\cr \end{array}\]
 and they have the probabilities $c_1=.23,\ c_2=.34,\ c_3=.43$.
\begin{figure}
\includegraphics[height = 4in,width = 6in]{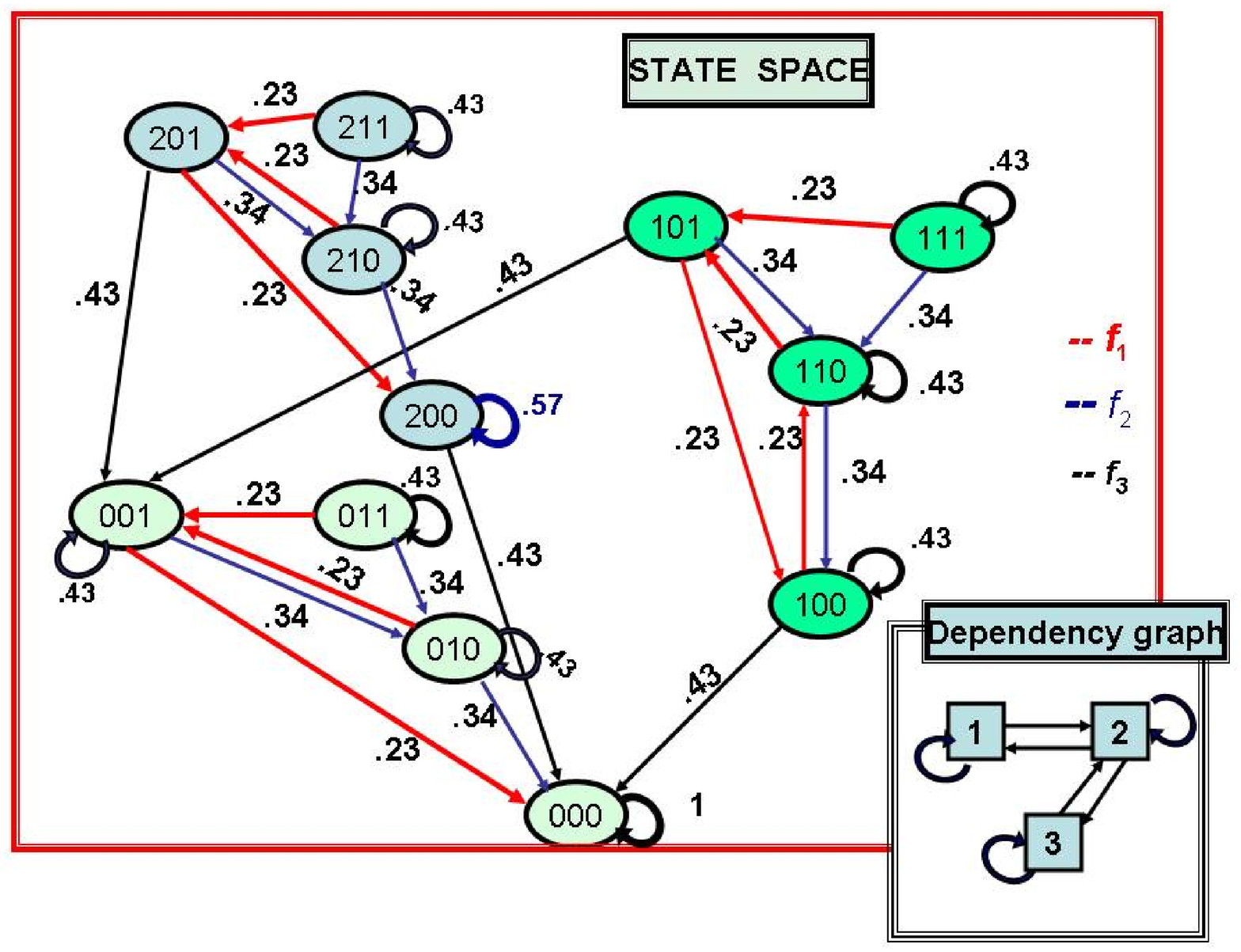}\caption{State Space of $\mathcal{G}$}
\end{figure}

The state space of $\mathcal{G}=(X,F,C)$ is in Figure $1$.  The
network has $12$ states.The only fixed point is $(0,0,0)$, and the
state space has two subnetworks of 8 elements and one subnetwork of
4 elements. For each subnetwork we must have a projection. That is,
an $\epsilon$-homomorphism $\pi:X\rightarrow Y$, must exist for each
subnetwork $Y$. That is, the converse of the Theorem \ref{theo}
could be true in some cases or with some little changes.

 In particular, for the sub-PRN $\mathcal{Y}_1=\{Y_1;F;C\}$ with
 \[Y_1=\{(1,0,0),(1,0,1),(1,1,0),(1,1,1),(0,0,0),(0,0,1),(0,1,0),(0,1,1)\},\]
  a projection $\pi_1:X\rightarrow Y_1$ exists, in fact: $\pi_1(x1,x2,x3)=(x1,x2,x3)$ if $x1=0,1$; and
$\pi_1(x1,x2,x3)=(0,x2,x3)$ if $x1=2$.  With this projection, it is
possible to consider the first gene with only two values: $\{0,
1\}$.

 For the sub-PRN $\mathcal{Y}_2=\{Y_2;F;C\}$ with
\[Y_2=\{(2,0,0),(2,0,1),(2,1,0),(2,1,1),(0,0,0),(0,0,1),(0,1,0),(0,1,1)\},\]
 a projection $\pi_2: X\rightarrow Y_2$ doesn't exist, because the first function $f_1$. So, taking a subnetwork of the whole PRN
 but without the function $f_1$ and a new assignation of probabilities we have a new PRN
 $\overline {\mathcal{G}}=\{X; f_2,f_3; d_2,d_3\}$ and a projection $\pi_2:\overline {\mathcal{G}}\rightarrow \overline {\mathcal{Y}_2}$ exists,
 and  it is given by: $\pi_2(x1,x2,x3)=(x1,x2,x3)$ if $x1=0,2$; and
$\pi_2(x1,x2,x3)=(0,x2,x3)$ if  $x1=1$, where $\overline
{\mathcal{Y}}_2=\{Y_2; f_2,f_3; d_2,d_3\}$ The projections $\pi_i$
are $.57$-homomorphisms. These two subnetworks $\overline
{\mathcal{Y}}_1$ and $\overline {\mathcal{Y}_2}$ are not  similar.
\begin{figure}
\includegraphics[height = 3in,width = 6in]{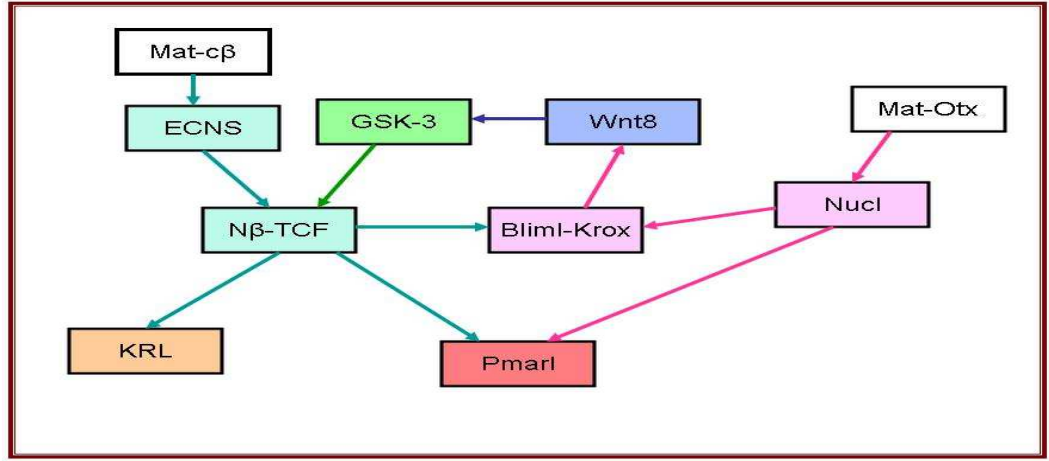}\caption{ Mat-Act Dependency graph}
\end{figure}

\subsection{Future work} The construction of  a mathematical model
for the genetic regulatory network \emph{ENDOMESODERM GENE NETWORK},
described in \cite{D,BT}, will be developed in the future. The
subnetwork, ``Mat-Act" is formed with the action of two genes
called: Mat-c$\beta$ and Mat-Otx over eight genes: ECNS, GSK3, Wnt8,
N$\beta$-TCF, Bliml-Krox, Nucl, KRL, Pmarl; whose interaction is
during 21 hours.  We will use the above methodology, for the genetic
network with the dependency graph  in Figure 2, obtained in
Biotapestry \cite{BT}.
\section{Acknowledgements}
This research was supported by the National Institute of Health,
 PROGRAM SCORE, 2004-08, 546112, University of Puerto Rico-Rio Piedras Campus, IDEA Network of Biomedical
  Research Excellence, and the Laboratory Gauss University of Puerto Rico Research.
   The first author wants to thank Professor E. Dougherty for his useful suggestions, and Professor O. Moreno for  his support
   during the last four years.

\end{document}